\documentclass[12pt,preprint]{aastex}

\newcommand{\simgt}{\lower.5ex\hbox{$\; \buildrel > \over \sim \;$}}
\newcommand{\simlt}{\lower.5ex\hbox{$\; \buildrel < \over \sim \;$}}

\newcommand{\sbkt}[1]{\left(#1\right)}
\newcommand{\abs}[1]{\left|#1\right|}

\shorttitle{}
\shortauthors{}
 
\begin{document}
 
\title{Observing H$_2$ Emission in Forming Galaxies}
\author{Kazuyuki Omukai \altaffilmark{1,2}
and Tetsu Kitayama \altaffilmark{3}}
\altaffiltext{1}{Department of Physics, Denys Wilkinson Building, 
Keble Road, Oxford, OX1 3RH, UK}
\altaffiltext{2}{Division of Theoretical Astrophysics, 
National Astronomical Observatory, Mitaka, Tokyo 181-8588, Japan}
\altaffiltext{3}{Department of Physics, Toho University, 
Funabashi, Chiba 274-8510, Japan}
\email{omukai@astro.ox.ac.uk; kitayama@ph.sci.toho-u.ac.jp}

\begin{abstract}
We study the H$_2$ cooling emission of forming galaxies, and discuss
their observability using the future infrared facility {\it SAFIR}.  Forming
galaxies with mass $\la 10^{11}M_{\sun}$ emit most of their
gravitational energy liberated by contraction in molecular hydrogen line
radiation, although a large part of thermal energy at virialization is
radiated away by the H Ly$\alpha$ emission.  For more massive objects,
the degree of heating due to dissipation of kinetic energy is so great that the
temperature does not drop below $10^{4}$K and the gravitational energy
is emitted mainly by the Ly$\alpha$ emission.  Therefore, the total
H$_2$ luminosity attains the peak value of $L_{\rm H_2} \sim 10^{42}
{\rm ergs/s}$ for forming galaxies whose total mass $M_{\rm tot} \sim
10^{11}M_{\sun}$.  If these sources are situated at redshift $z \sim 8$, they
can be detected by rotational lines of 0-0S(3) at 9.7$\mu$m and 0-0S(1)
at 17$\mu$m by {\it SAFIR}.  An efficient way to find such H$_2$
emitters is to look at the Ly$\alpha$ emitters, since the brightest
H$_2$ emitters are also luminous in the Ly$\alpha$ emission.
\end{abstract}
 
 
\section{Introduction}
In the last decade, observational studies on primeval galaxies 
has made great progress.
For example, deep surveys in the optical (Williams et al. 1996), 
near-infrared (Maihara et al. 2001) and 
mid-infrared (Taniguchi et al. 1997) wavelengths 
have revealed early evolution of galaxies.

Since galaxies in the earliest phase consist mostly of gas, 
knowledge on interstellar medium in forming galaxies 
is valuable for understanding galaxy formation process.
Molecular line observation has been a powerful 
tool in diagnosing the physical conditions in local star-forming regions 
(e.g., Dyson \& Williams 1997).  Recently, molecular line emission 
has also been found to emanate from high-redshift galaxies
and the number of detected sources has been increasing rapidly
(e.g., Ohta et al. 1996; Omont et al. 1996; Combes 1999).  
Among theoretical studies, Spaans \& Silk (1997)
investigated CO line emission from protogalaxies.  However, the 
detection of metal lines depends on the degree of metal enrichment 
and thus preceeding star formation is required.

Can we detect molecular lines in primordial pristine gas clouds?  The
possibility of detecting H$_2$ line emission from forming galaxies was
discussed as early as the firm establishment of the Big-Bang cosmology
(Saslaw \& Zipoy 1967).  Other pioneering work includes that by Izotov
\& Kolesnik (1984) and Shchekinov \& \'{E}nt\'{e}l' (1985), which treats
the H$_2$ emission from proto-clusters of galaxies, i.e, the so-called
``Zel'dovich pancakes''.  More recently, Ciardi \& Ferrara (2001)
investigated the H$_2$ emission from the super-shells created in the
intergalactic medium as a result of multiple SN explosions in PopIII
objects.  They concluded that about 10 per cent of the SN mechanical
energy is carried away by the H$_2$ 1-0S(1) rovibrational line.  
Stimulated by the recent theoretical understanding
of the first star-forming regions (e.g., Omukai \& Nishi 1998; Abel,
Bryan, \& Norman 2002; Bromm, Coppi, \& Larson 2002; Omukai \& Palla
2001, 2003), research has been vigorously carried out on 
the radiation from individual star-forming cores.  
Flower \& Pineau des For\^{e}ts (2001) calculated
H$_2$ emission from a J-type shock expected to appear in the prestellar
collapse.  Kamaya \& Silk (2002) studied the H$_2$ emission from a
collapsing prestellar core using a simplified model, and pointed out
that the 0-0S(3) line is the strongest during the prestellar collapse phase.
Ripamonti et al. (2002) arrived at similar conclusions by a more refined
calculation.  Shibai et al. (2001) studied the detectability of H$_2$ 
absorption lines in high-redshift objects.  All these studies,
however, suggest that the prospect for direct detection of H$_2$
emission from forming galaxies is rather dim, though not rejected, even
with next generation facilities. 

On the other hand, Ly$\alpha$ emitting galaxies in high-redshifts
have been observed (e.g., Hu, Cowie, \& McMahon 1998; Steidel et
al. 2000; Kodaira et al. 2003).  
Stimulated by these findings, some authors have investigated the
Ly$\alpha$ cooling radiation from protogalaxies by semi-analytical
modeling (Haiman, Spaans, \& Quataert 2000) or by numerical simulations
(Fardal et al. 2001).  They concluded that the greater part of the
gravitational energy is indeed emitted by the H Ly$\alpha$ line emission.  In
their model, however, H$_2$-line cooling is not included, and the
temperature in protogalaxies thus remains above $10^4$ K, below which
atomic cooling is strongly suppressed.  Although this is a fair assumption
for the massive objects that they considered (see \S 3 below),
for smaller objects with total mass $<10^{11}M_{\sun}$, sufficient H$_2$
can be formed as a result of enhanced ionization degree by virialization
shock and its delayed non-equilibrium recombination (Shapiro \& Kang
1986; Susa et al. 1998; Oh \& Haiman 2002).  By this mechanism, the
temperature drops further below $10^4$ K due to H$_2$ line cooling.
Therefore, the greater part of gravitational energy liberated by
contraction will be available for the H$_2$ cooling radiation.  In this
paper, we study the H$_2$ cooling radiation from those forming galaxies
and discuss the prospect for observing such sources.  Our result is
quite encouraging, suggesting that H$_2$ emission is indeed observable
by the future infrared satellite mission ${\it SAFIR}$ from sources as
high as $z=8$.

The organization of the rest of this paper is as follows:  In \S 2, the
method of calculation is described.  In \S 3, the prediction for the
H$_2$ emission from forming galaxies and its observability is discussed.
Finally in \S4, a brief summary and discussions are presented.
Throughout the paper, we adopt a $\Lambda$CDM cosmology in accordance
with the WMAP data (Spergel et al. 2003): $(\Omega_{\rm
M},\Omega_\Lambda , \Omega_{\rm B}, h, \sigma_8)
=(0.3, 0.7, 0.05, 0.7, 0.9)$.

\section{Model}

\subsection{H$_2$ emission from forming galaxies}
\label{secemit}

In this section, we describe our model for evaluating H$_2$ line
emission.  We use a simplified evolution model where a forming
protogalaxy is treated as a uniform cloud.

As the initial state, we take an instance of virialization.
We use the analytic formula for the density at virialization 
\begin{equation}
\rho_{\rm vir}=\rho_{\rm crit} \Delta_{\rm c}
\end{equation}
where
\begin{equation}
\Delta_{\rm c}=18 \pi^2 +82x -39x^2,~~ 
x=-\frac{\Omega_{\Lambda}}
{\Omega_{0}(1+z_{\rm vir})^3 + \Omega_{\Lambda}},
\end{equation}
which has been found to provide a good description 
of the numerical results (Bryan \& Norman 1998).

The critical density at $z=z_{\rm vir}$ can be written as
\begin{equation}
\rho_{\rm crit}=\rho_{\rm crit, 0}
[\Omega_0(1+z_{\rm vir})^3+\Omega_{\Lambda}],
\end{equation}
where $\rho_{\rm crit, 0}$ is the present-day critical density.

The radius and temperature of objects of total mass $M_{\rm tot}$ 
in this instance are
\begin{equation}
R_{\rm vir}=\left( M_{\rm tot} / \frac{4}{3} \pi \rho_{\rm vir} \right)^{1/3},
\end{equation}
and
\begin{equation}
T_{\rm vir}=\frac{G \mu m_{\rm H}}{5 k_{\rm B} R_{\rm vir}} M_{\rm tot},
\end{equation}
respectively.
 
We first describe the dissipationless case and later describe the
prescription for the dissipation of kinetic energy.  For the sake of
simplicity, the dark matter density is assumed to be constant after the
virialization.  The evolution of gas density is computed by solving the
virial equation for a spherical object
\begin{equation}
\frac{d^2 R}{dt^2}=-\frac{G M}{R^2} \left(1- \frac{T}{T_{\rm vir}(R)} \right)
\label{eqmotionf0}
\end{equation}
where
\begin{equation}
T_{\rm vir}(R)=\frac{G \mu m_{\rm H}}{5 k_{\rm B} R} M(R),
\end{equation}
and $M(R)$ is the mass enclosed within a spherical shell of radius $R$.

The thermal evolution is computed by solving the energy equation
\begin{equation}
\frac{de}{dt}=-p\frac{d}{dt} \left(\frac{1}{\rho} \right)
- \Lambda_{\rm rad} - \Lambda_{\rm chem},
\label{eqenergyf0}
\end{equation}
where $e$ is the specific thermal energy
\begin{equation}
e=\frac{1}{\gamma_{\rm ad} -1} \frac{k_{\rm B} T}{\mu m_{\rm H}},
\end{equation}
and $\Lambda_{\rm rad}$ and $\Lambda_{\rm chem}$ are the radiative and
chemical cooling rates per unit mass, respectively.

The radiative cooling rate includes H$_2$ line emission, HD line
emission, atomic emission by H and He, and Compton cooling.  The H$_2$
line cooling is treated as in Omukai (2001), where the effect of CMB is
taken into account and the ortho-para ratio of 3:1 is assumed.  For the
HD cooling, the cooling function by Galli \& Palla (1998) is used.  The
effect of CMB for HD cooling is treated approximately by replacing the
cooling rate by HD line  $\Lambda_{\rm HD}(T)$ by 
$\Lambda_{\rm HD}(T)-\Lambda_{\rm HD}(T_{\rm rad})$
(e.g., Tegmark et al. 1997). The atomic emission includes collisional 
excitation cooling, recombination cooling,
and bremsstrahlung of H and He.  The coefficients given by Anninos et
al. (1999) are used.

The above formulation neglects dissipation of kinetic energy of
contracting gas; after the initial thermal energy is carried away by
radiation, essentially all the gravitational energy liberated by
contraction contributes to accelerating the gas infall.  If the gas,
however, settles into a Kepler rotation disk, the kinetic energy is
reduced by a factor of 2 compared to the free-falling sphere with the
same radius, i.e. half of the kinetic energy must be dissipated away.
Such dissipation may proceed through the thermalization of kinetic
energy by turbulent motion induced by the non-spherical nature of the
collapse.  To mimic this effect, we add a dissipation term in equations
of motion and of energy;
\begin{equation}
\frac{d^2 R}{dt^2}=-\frac{G M}{R^2} \left(1- \frac{T}{T_{\rm vir}(R)} \right)
+D_{\rm trb}
\label{eqmotionf025}
\end{equation}
and
\begin{equation}
\frac{de}{dt}=-p\frac{d}{dt} \left(\frac{1}{\rho} \right)
- \Lambda_{\rm rad} - \Lambda_{\rm chem}+ \Gamma_{\rm trb},
\label{eqenergyf025}
\end{equation}
where $D_{\rm trb}$ and $\Gamma_{\rm trb}$ are the drag and heating
terms due to dissipation of kinetic energy. To compute these quantities,
we assume that a fraction $f_{\rm trb}$ of kinetic energy is
thermalized during the collapse. Specifically, we first evaluate at each
timestep the increment of kinetic energy in the dissipationless case
$\Delta K_{0}$ using equations (\ref{eqmotionf0}) and
(\ref{eqenergyf0}).  We then solve equations (\ref{eqmotionf025}) and
(\ref{eqenergyf025}) assuming that $f_{\rm trb} \Delta K_{0} =
\Gamma_{\rm turb} \Delta t$ is injected as heat and the actual increment
of kinetic energy is $\Delta K=(1-f_{\rm trb}) \Delta K_{0}$.  In
reality, the value of $f_{\rm trb}$ is likely to lie between $0.5$ (for
the Kepler rotating disk) and $0$ (for the spherical collapse).  We
adopt $f_{\rm trb}=0.25$ as our fiducial value, while we also study the
dissipationless case ($f_{\rm trb}=0$) for comparison.

The freeze-out values of chemical abundances are used for the 
initial abundances for the computation: $y_{e}=3\times 10^{-4}, 
y_{H_{2}}=1 \times 10^{-6}$ and $y_{HD}=1 \times 10^{-9}$ 
(Galli \& Palla 1998). 
The chemical reactions between H, and He compounds are solved following 
Omukai (2000; reactions 1-22), with HD-bearing reactions of   
D3, D4, D8,and D10 in Galli \& Palla (1998) added. 
We find, however, that HD cooling is not important in all cases
we studied.

The cloud collapse is followed from the virialization until 
either the local Hubble time
\begin{equation}
t_{\rm H}(z)=\frac{2}{3H_{0}}(1-\Omega_{0})^{-1/2} {\rm sinh}^{-1}
\left[\frac{(1-\Omega_{0})^{1/2}}{\Omega_{0}^{1/2}(1+z)^{3/2}}\right]
\end{equation}
elapses or the radius reaches the rotation barrier
\begin{equation}
R_{\rm rot}=0.06 \sbkt{\frac{f_{\rm gas}}{0.17}}^{-1} 
\sbkt{\frac{\lambda_{\rm turn}}{0.05}}^2 R_{\rm vir}, 
\end{equation}
where $f_{\rm gas}$ is the mass fraction of gas.
We adopt a median value 0.05 for the dimensionless spin parameter at
turn-around $\lambda_{\rm turn}$ (Barns \& Efstathiou 1987) and the gas
density enhances by a factor of $5000$ between $R_{\rm vir}$ and $R_{\rm
rot}$.  The duration of emission is therefore
\begin{equation}
t_{\rm em}(M,z) = \min [t_{\rm H}(z), t_{\rm rot}(M,z)],  
\label{eq_teq}
\end{equation}
where $t_{\rm rot}(M,z)$ is the time taken to reach $R_{\rm rot}$ from
$R_{\rm vir}$.

\subsection{Merger probability of protogalaxies}

In predicting the luminosity and the abundance of protogalaxies, we also
take into account explicitly hierarchical clustering and mergers of
these objects.  This is of particular importance at high redshift, say
$z \simgt 8$, where a significant fraction of galaxy-scale halos merges
into larger halos before the free-fall time is reached. For such halos,
the duration of emission will be shorter than that estimated in 
Sec \ref{secemit}.

We incorporate the variations in the emission times by defining the
``merger probability weighed average luminosity'' of a protogalaxy with
total mass $M$ at redshift $z$ as
\begin{eqnarray}
L_{\rm line}(M,z) =\frac{ 
\int_{t(z)-t_{\rm em}(M,z)}^{t(z)}
L_{\rm line}^{\rm int}(M,z,z_f) \frac{dp(M,z,z_f)}{dz_f} 
\abs{\frac{dz_f}{dt_f}} dt_f}{
\int_{t(z)-t_{\rm em}(M,z)}^{t(z)}
\frac{dp(M,z,z_f)}{dz_f} \abs{\frac{dz_f}{dt_f}} dt_f}
\label{eqlum}
\end{eqnarray}
where $L_{\rm line}^{\rm int}(M,z,z_f)$ is the intrinsic line luminosity
at $z$ of the galaxy virialized at $z_f$, corresponding to the cosmic
time $t_f \equiv t(z_f)$, and $dp(M,z,z_f)/dz_f$ is the probability
distribution of formation epochs of virialized halos.  The former
quantity is computed using the model described in Sec \ref{secemit}.
Since $t_{\rm em}(M,z)$ is typically an order of magnitude smaller than
$t(z)$, we neglect the changes in $\rho_{\rm vir}$ and $R_{\rm vir}$
between $z$ and $z_f$.  For the latter we adopt the prescription of
Lacey \& Cole (1993) as described in Kitayama \& Suto (1996); the
formation epoch $z_f$ is defined as the time at which a halo assembles
at least half of its final mass. In the rest of this paper, unless
otherwise stated, we simply refer as ``luminosity'' to that given in
equation (\ref{eqlum}).

The number density of protogalaxies in accord with the above definition
of average luminosity is 
\begin{equation}
\frac{dn(M,z; L_{\rm line})}{dM} = 
\frac{dn_{\rm PS}(M,z)}{dM}  \int_{t(z)-t_{\rm em}(M,z)}^{t(z)}
\frac{dp(M,z,z_f)}{dz_f} \abs{\frac{dz_f}{dt_f}} dt_f,
\end{equation}
where $dn_{\rm PS}(M,z)/dM$ is the Press-Schechter (1974) mass function.

If strong bias is present in such high redshift we consider here, 
namely $z \sim 10-20$, the number of observable objects increases 
remarkably. In this sense, our result, which corresponds to the no 
bias case, can be regarded as the lower bound on the number of objects.

\section{Results}
\subsection{H$_2$ emission from individual objects}

In the following, we consider two epochs in the early universe,  
$z=$20 and 8.
The former corresponds to the typical epoch for the first star
formation (e.g., Abel et al. 2002).
The latter is believed to be shortly before the completion 
of the cosmic reionization from the observation of high-redshift
quasars (e.g., Becker et al. 2001).
Direct observation of objects forming in this epoch is therefore 
intersting for understanding the reionization process.

The total luminosity in H$_2$ lines is shown in Figure 1 as a function
of the total mass of the objects, along with the Ly$\alpha$ luminosity
for the virial redshift (a)$z_{\rm vir}=$20 and (b)$z_{\rm vir}=$8.  
The corresponding virial temperature and the observed flux are indicated in
the top and right axes.  The observed flux in a line is calculated using
the usual relation
\begin{equation}
S_{\rm line}=L_{\rm line}/4 \pi d_{L}(z)^2,
\end{equation}
where $d_{L}(z)$ is the luminosity distance to the source at redshift
$z$.  The contribution to the total H$_2$ luminosity from each line is
shown in Figure 2 for the fiducial case ($f_{\rm trb}=0.25$) and in
Figure 3 for the dissipationless case ($f_{\rm trb}=0$).
Important lines are listed in Table 1 (for $z=20$)
and 2 (for $z=8$) for the fiducial case.

We first focus on the fiducial case $f_{\rm trb}=0.25$.  From the H$_2$
emission property, the protogalactic clouds can be categorized into the 
following four mass ranges.  The temperature evolution for typical cases 
in these regimes is shown in Figure 4 for $z=20$.
\begin{itemize}
\item {\it Pop III objects}~~($M_{\rm tot} \la 10^{8}M_{\sun}$)

Small objects whose virial temperature $T_{\rm vir}\la 10^{4}$K cannot
excite atomic hydrogen, and cool only by H$_2$ lines.  Thus both the
thermal energy at virialization and the gravitational energy liberated
by the collapse are converted to H$_2$ emission.  The dominant lines are
the lowest excitation ones, e.g., 0-0S(0) at 28 $\mu$m and 0-0S(1) at 
17 $\mu$m (see Figure 2).

\item {\it dwarf galaxies}~~ 
($10^{8}M_{\sun} \la M_{\rm tot} \la 10^{11}M_{\sun}$)

Clouds with a virial temperature exceeding $10^4$K can cool by 
Ly$\alpha$ radiation.
As we can see in Figure 4, the temperature first drops rapidly from 
the virial temperature to that appropriate for H$_2$ cooling 
($\sim 10^{3}$K).
The initial thermal energy is radiated away by the atomic radiation 
processes including the Ly$\alpha$ emission ($T \ga 10^{4}$K)
and by the H$_2$ line emission ($T \la 10^{4}$K).
The collapse proceeds quasi-isothermally thereafter   
due to the H$_2$ line emission.
This can be observed in Figure 1 as the emergence of Ly$\alpha$ luminosity.
At the same time, the slope of the increase of the H$_2$ line luminosity 
toward higher mass flattens slightly around $T_{\rm vir} \simeq 10^{4}$K. 
The low excitation lines, e.g. 0-0S(1) at 17 $\mu$m and 0-0S(3) at 
9.7 $\mu$m, are still dominant (Figure 2).

\item {\it giant galaxies}~~
($10^{11}M_{\sun} \la M_{\rm tot} \la 10^{12}M_{\sun}$)

Since the available gravitational energy increases toward massive objects, 
the heating rate by the dissipation of kinetic energy also increases.
For objects $> 10^{11}M_{\sun}$, the heating becomes so great that 
the temperature does not fall below $10^{4}$K.
These objects collapse isothermally at $10^{4}$K, 
after initial rapid cooling (see Figure 4 a).
During this isothermal collapse, a large fraction of energy is emitted in
the Ly$\alpha$ radiation.
Then the Ly$\alpha$ luminosity exceeds the H$_2$ luminosity (Figure 1) for
those objects.
Despite the high temperature, a small amount of H$_2$ is present.
Only a small fraction of gravitational energy is emitted in the H$_2$ lines.
This can be observed as a plateau in the H$_2$ luminosity in Figure 1, 
as the larger mass compensates for the lower H$_2$ emissivity.
Corresponding to the abrupt change in thermal property, the dominant lines 
are altered (see Figure 2).  
Because the collapse proceeds with high temperature, high excitation 
lines become more important.
The strongest line is 0-0S(3) at $9.7 \mu$m.
Although 0-0S(1) is still in the second place, the vibrational line 
1-0S(3) at $2.0 \mu$m also becomes important. 
These objects are the brightest ones in H$_2$ emission and 
the H$_2$ luminosity reaches about $3 \times 10^{41}{\rm ergs~s^{-1}}$.
Note that these are also most luminous in Ly$\alpha$ emission.

\item {\it no-cooling objects} ($M_{\rm tot}\ga 10^{12}M_{\sun}$)

For these massive objects, 
both Ly$\alpha$ and H$_2$ luminosities decline rapidly,
because, in the redshift range we are considering, 
such massive objects cannot cool within the merging 
timescale and then remain in the virialization state (Silk 1977; 
Rees \& Ostriker 1977).
\end{itemize}

We stress the importance of the dissipation of kinetic energy.
Early studies by Saslow \& Zipoy (1967) and by Izotov \& Kolesnik (1984)
did not take the dissipation into account.  As we can see in Figure 1,
the maximum H$_2$ luminosity would be smaller by an order of magnitude
without the dissipation of kinematic energy.  In this case, only the
portion of the initial thermal energy below 10$^4$K would be available
for H$_2$ emission, with most gravitational energy being converted into
kinetic energy.  The discrepancy in emitted energy increases toward
higher mass because of the larger gravitational energy to be released 
during the collapse; in the dissipationless case, the H$_2$ luminosity 
increases only linearly in mass, while in the dissipational case it increases
approximately as $\propto M_{\rm tot}^{1.5}$.

\subsection{Number counts of H$_2$ emitters and their contribution 
to the far-IR background light}

H$_2$-line photons emitted at $z\simeq 10$ have been red-shifted to the
far-infrared (FIR) wavelength today.  The future space telescope, the
{\it Single Aperture Far-Infrared
Observatory}\footnote{http://safir.jpl.nasa.gov/index.asp} ({\it
SAFIR}), is planned to have unprecedented sensitivity at the FIR and
submillimeter wavelengths, with the limiting flux of $\sim 10^{-19} {\rm
erg~s^{-1} cm^{-2}}$ in FIR.  With this limiting
flux, brightest sources at $z=8$ can be observable by the 0-0S(3) 9.7$\mu$m
line (Figure 2 b), while at redshift $z=20$, even the brightest H$_2$ emitters 
are below this threshold (Figure 2 a).

Figure 5 shows the number of sources per unit solid angle above the
limiting flux of (a) $10^{-19} {\rm erg ~s^{-1} cm^{-2}}$ and (b)
$10^{-20} {\rm erg ~s^{-1} cm^{-2}}$ whose emission redshift is larger
than $z$.  For the limiting flux $10^{-19} {\rm erg ~s^{-1} cm^{-2}}$,
which is the planned limit for {\it SAFIR}, although the number of
sources with their H$_2$ total flux exceeding this limit is about
10$^{6} ~{\rm str^{-1}}$ as early as $z=10$, even the strongest line
0-0S(3) 9.7 $\mu$m cannot be observed until $z=8$ (see Figure 5 a).  We
can understand this from Figure 2 (b); at $z=8$, although the maximum
total H$_2$ flux exceeds $10^{-19}{\rm erg~s^{-1} cm^{-2}}$ by some
factor, even the most dominant line 0-0S(3) 9.7$\mu$m barely reaches
this limiting flux because of sharing the luminosity with other lines.

For the limiting flux $10^{-20} {\rm erg ~s^{-1} cm^{-2}}$, the
brightest objects at $z=20$ could be visible by the 0-0S(3) 9.7$\mu$m
line and the 0-0S(1) 17$\mu$m line (Figure 2 a), although such massive
objects are still very rare at $z=20$ (Figure 5 b). For this limiting
flux, the number of sources observable by these lines exceeds $\sim 100$
str$^{-1}$ at $z \simlt 15$.

Figure 6 further illustrates the number of H$_2$ emitters for each line
above the limiting flux per unit redshift range at $z=8$.  At the
limiting flux of {\it SAFIR}, $10^{-19}{\rm erg~s^{-1} cm^{-2}}$,
observable lines are 0-0S(3) 9.7$\mu$m, and possibly 0-0S(1) 17$\mu$ m,
although the latter slightly falls short of $10^{-19}{\rm erg~s^{-1}
cm^{-2}}$.  If the limiting flux were an order of magnitude deeper,
some vibrational lines including 1-0Q(1) 2.4 $\mu$m and 1-0S(1) 2.1$\mu$m 
could also be detected.

Finally, Figure 7 displays the contribution of these H$_2$ emitters 
to the extragalactic FIR background radiation.  Also plotted for
reference is the FIR background spectrum detected by {\it COBE/FIRAS}
(Fixsen et al. 1998).  The H$_2$ line emission from forming galaxies
makes only a negligible contribution to the background light, being smaller
than the {\it COBE/FIRAS} result by several orders of magnitude.

\section{Summary and Discussion}
We have evaluated the amount of H$_2$ line radiation from forming
galaxies.  The H$_2$ line luminosity reaches the maximum value of about
$3 \times 10^{41} {\rm erg~s^{-1}}$ for protogalaxies of total mass 
$M_{\rm tot}\sim 10^{11}M_{\sun}$.  
Below this mass scale, all the gravitational energy
is emitted by H$_2$ line radiation, while above this scale, a large part
of the energy is emitted by Ly$\alpha$ radiation.  Such brightest objects at
$z \sim 8$ are detectable by the next generation telescope {\it SAFIR}
via the 0-0S(3) 9.7$\mu$m line emission and marginally detectable also
via the 0-0S(1)17$\mu$m line.  The cosmic FIR background radiation by
the H$_2$ emission is several orders of magnitude below the detected
value of {\it COBE/FIRAS}, and can be safely neglected.

The H$_2$ luminosity from a protogalactic cloud has also been studied 
by Shchekinov (1991).
Assuming that a protogalactic cloud consists of subclumps
and that the luminosity is given by the sum of luminosity from 
each subclumps, he evaluated $L_{\rm H_2} \ga 10^{42} {\rm erg/sec}$ 
from a protogalactic cloud with gas mass $10^{11}M_{\sun}$ 
corresponding to total mass $M_{\rm tot} \simeq 10^{12}M_{\sun}$.
On the other hand, in our model 
the large protogalaxies of $M_{\rm tot} \ga 10^{12}M_{\sun}$ 
has no time to cool and then emit only weak H$_2$ radiation 
(total H$_2$ luminosty $L_{\rm H_2} \sim 10^{40} {\rm erg/sec}$). 
This is because such inhomogenous clouds are regarded as 
a cluster of protogalactic clouds rather than a single
cloud in our scheme. 
In fact, the sum of luminosity from 
a cluster with total mass of $M_{\rm tot} \simeq 10^{12}M_{\sun}$
gives a similar value as that of Shchekinov (1991).
For example, the luminosity of a protogalactic cloud of 
$M_{\rm tot}=10^{8}M_{\sun}$ collapsing at $z=20$ is 
$L_{\rm H_2} \sim 10^{8} {\rm erg/sec}$ (Fig. \ref{fig:f2} a).
Then the H$_2$ luminosity of a cluster consisting of 10$^{4}$ such 
clouds is $\sim 10^{12} {\rm erg/sec}$.
Note that in order for this value to be the actual luminosity 
from the cluster, the collapse of the subclumps must be simultaneous.
Since this is probably a rare occurence, this possibility is not considered 
in this paper.

As we mentioned in \S 3, the brightest H$_2$ sources, ``giant
galaxies'', are even more luminous in the Ly$\alpha$ emission.  
This has been pointed out also by Shchekinov (1991). 
who investigated pancaking shock of primordial clouds.
Consequently, the most
efficient way of finding H$_2$ emitters is to look at the already
detected Ly$\alpha$ emitters, rather than looking for them in a blank field.
The origin of Ly$\alpha$ emitters found at high-redshift is still in
dispute.  If Ly$\alpha$ emitters are originating in the cooling
radiation, they might be bright as well as in the H$_2$ emission.  On the
other hand, the Ly$\alpha$ emission might be originating in the
recombination radiation due to stellar ionization (Partridge \& Peebles
1967).  In this case, there is no reason to expect bright H$_2$ emission
from those sources.  Nonetheless, since H$_2$ is vulnerable to
photodissociation by stars as we will discuss below, even if 
the H$_2$ emission will not be detected, this will not demonstrate 
the recombination origin for the Ly$\alpha$ radiation.

Here, we discuss some effects that can alter the amount of the H$_2$
luminosity from forming galaxies. The density contrast inside the
clouds, in particular higher density than average in the inner region,
results in luminosity higher by some factor, reducing the collapse
timescale compared to the merging timescale.  On the other hand, the
following effects reduce the H$_2$ luminosity:  First, molecular
hydrogen is vulnerable to photodissociation by stars.  According to the
recent WMAP results, the reionization epoch of the universe could be as
early as $z\simeq 17 \pm 5$ (Spergel et al. 2003).  At least below this
redshift, protogalaxies are irradiated by the UV background radiation.
According to our study, objects most luminous in H$_2$ emission are
massive ones whose mass is about 10\% of $L_{\ast}$ galaxies today.
Such massive objects are not vulnerable to photodissociation by UV
background radiation.  Therefore, our results remain unchanged except
for ``PopIII objects'', for which the feedback effect is significant
(Haiman, Rees, \& Loeb 1997; Omukai \& Nishi 1999; 
Ciardi et al. 2000; Kitayama et al. 2001; Ricotti et al. 2002).  
Although ``giant galaxies'' are not affected by the UV background 
radiation, an internal UV field created
by local star formation might photodissociate molecular hydrogen and
then reduce the amount of the H$_2$ emission.  Another possibility for
reducing the H$_2$ emissivity is by metal enrichment; with metals, the
energy budget for H$_2$ emission is reduced by the cooling via metal
fine-structure and metastable lines (e.g., Dalgarno \& McCray 1972).  For
the metallicity $Z>10^{-2}Z_{\sun}$, the metal line cooling dominates
over the H$_2$ line cooling (e.g., Fall \& Rees 1985).  Below this
metallicity, the H$_2$ cooling dominates and thus the effect of metal lines
in reducing the H$_2$ luminosity can be neglected.  Further study of the
influence of stellar feedback and metal enrichment on the H$_2$ emission
will be interesting.

\acknowledgements 
We are grateful for helpful discussions with H. Matsuo, H. Susa, 
and M. Umemura.
This work is supported in part by the Research Fellowship
of the Japan Society for the Promotion of Science for Young Scientists
(6819, KO), and the Grants-in-Aid by the Ministry of Education,
Science and Culture of Japan (14740133, TK).
 
\appendix
\newpage


\newpage
\bigskip
\centerline{\bf Figures}

\plottwo{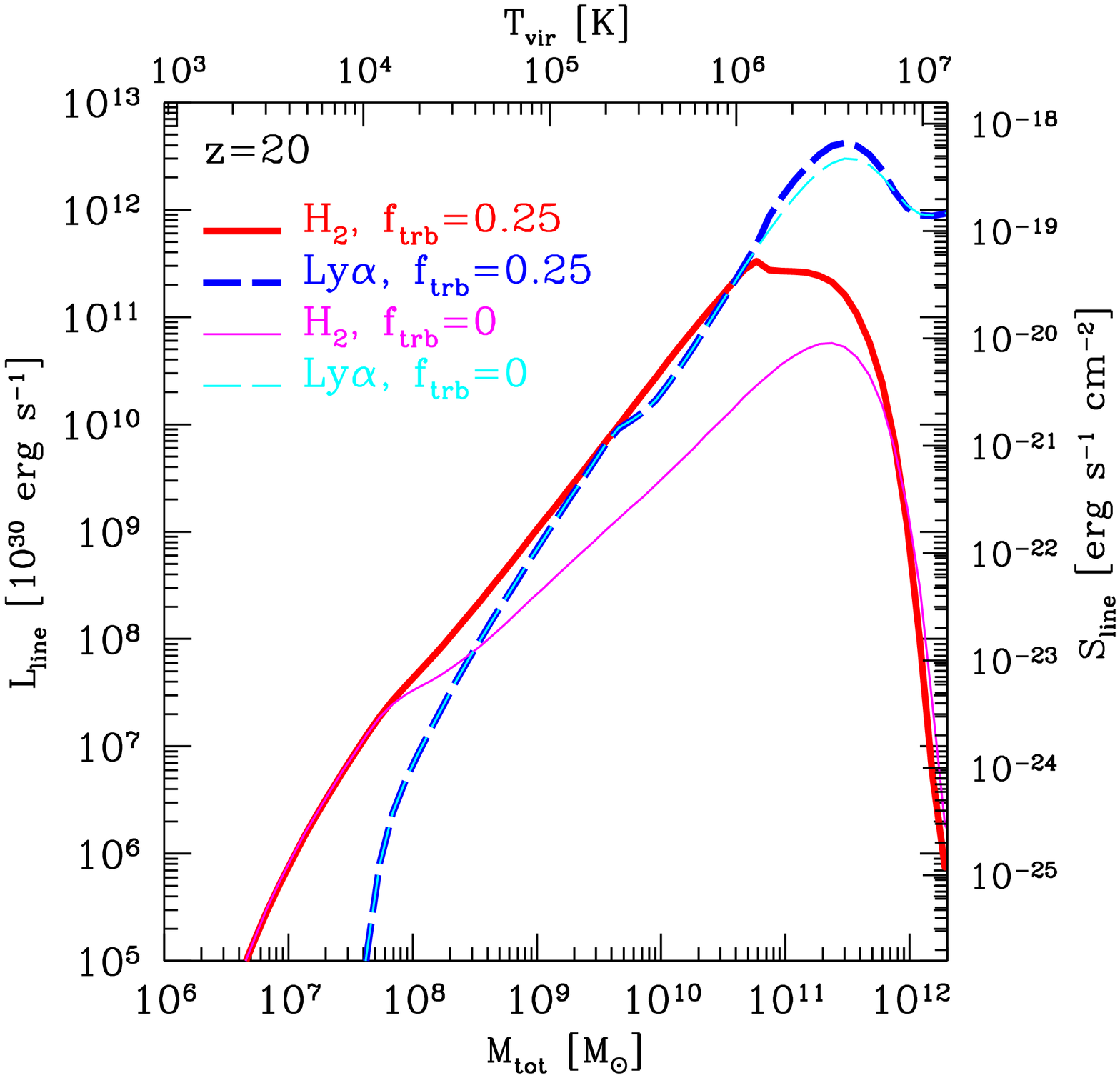}{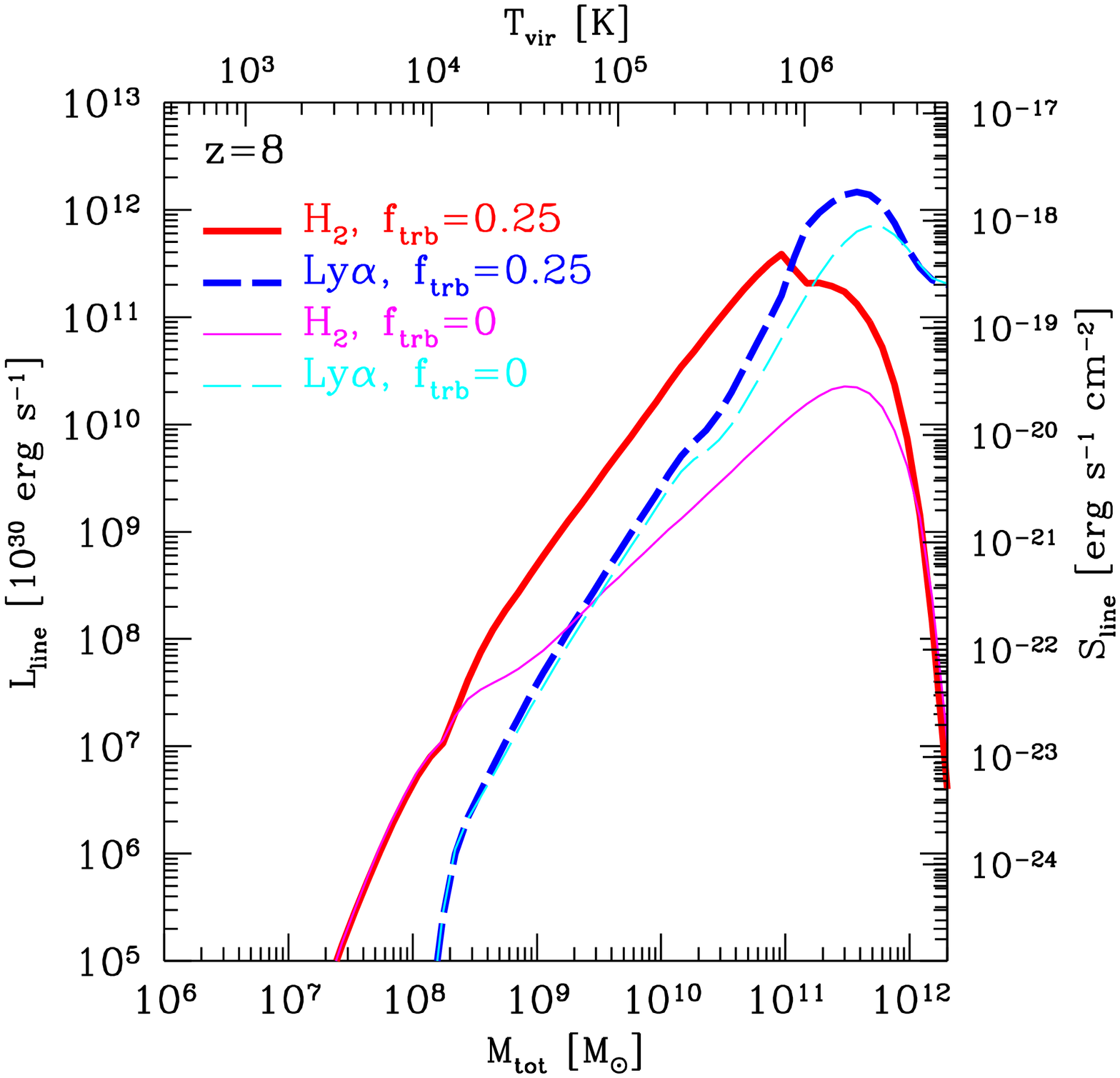}
\figcaption[f1a.eps;f1b.eps]{The total luminosity in H$_2$ lines and the
Ly$\alpha$ line luminosity as a function of the mass of protogalactic
clouds that collapses at (a) z=20, and (b) z=8.  The luminosity is
averaged over the merger probability of halos as in equation
(\ref{eqlum}). The corresponding virial temperature is also shown along
the top axis.  The observed flux today is shown along the right axis.
Both the fiducial case ($f_{\rm trb}=0.25$) and the dissipationless case
($f_{\rm trb}=0$) are shown.  \label{fig:f1}}
 
\plotone{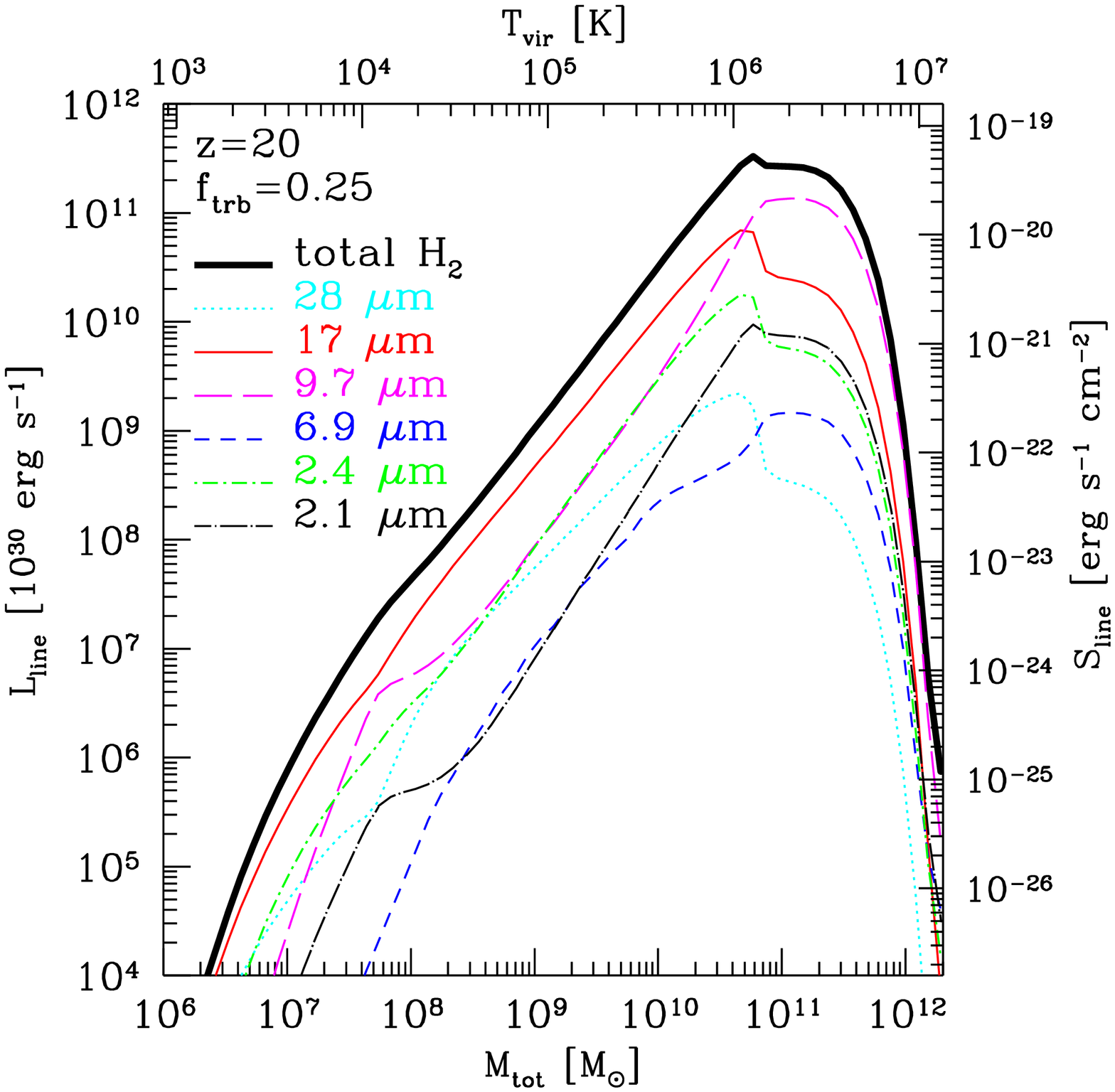}
\plotone{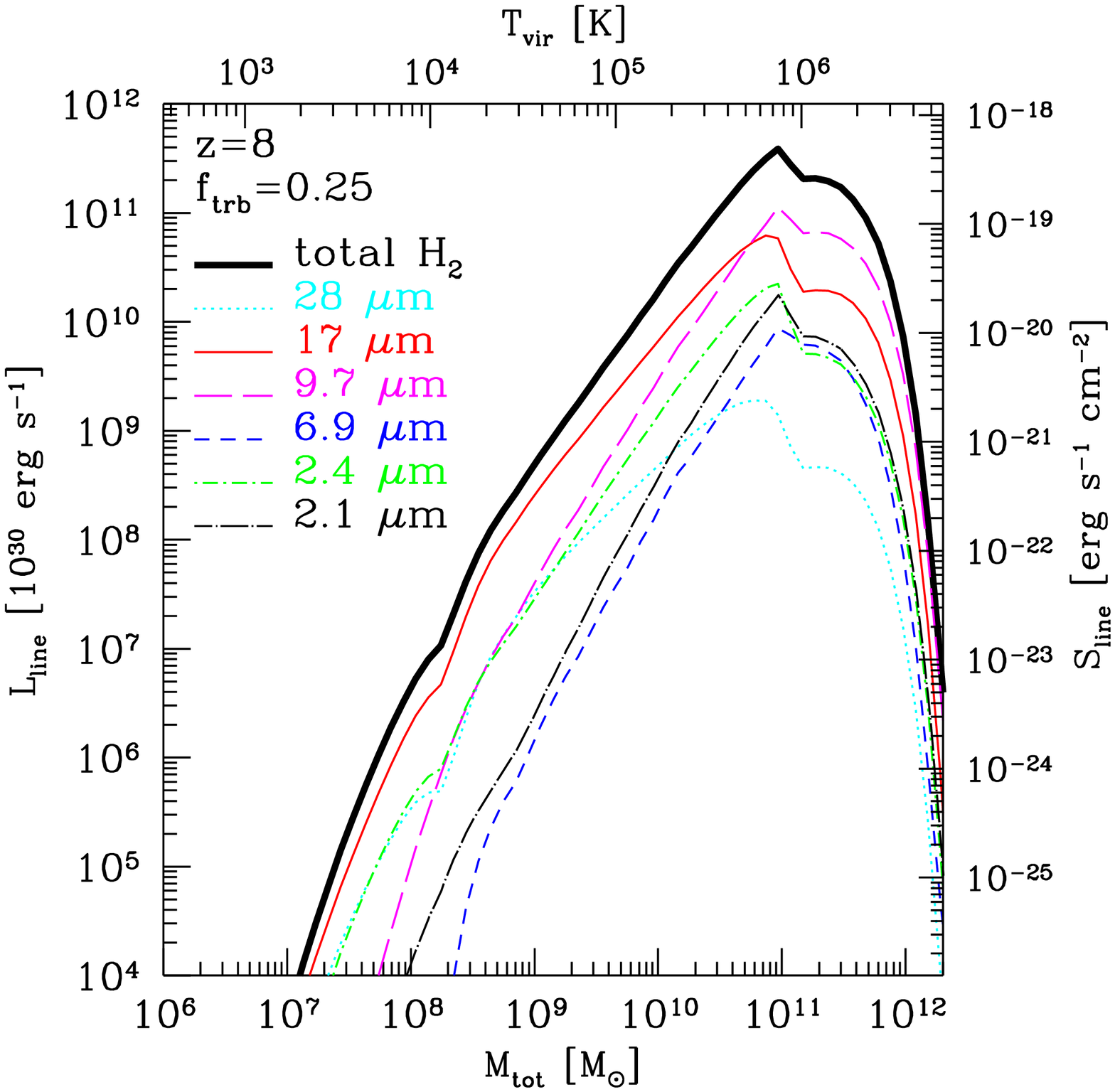}
\figcaption[f2a.eps;f2b.eps]{The contribution from individual lines to 
the H$_2$ line luminosity in the fiducial case ($f_{\rm trb}=0.25$) 
for the collapse redshift of (a) z=20 and (b) z=8. 
The thick solid line shows the sum of all lines.
\label{fig:f2}}

\plotone{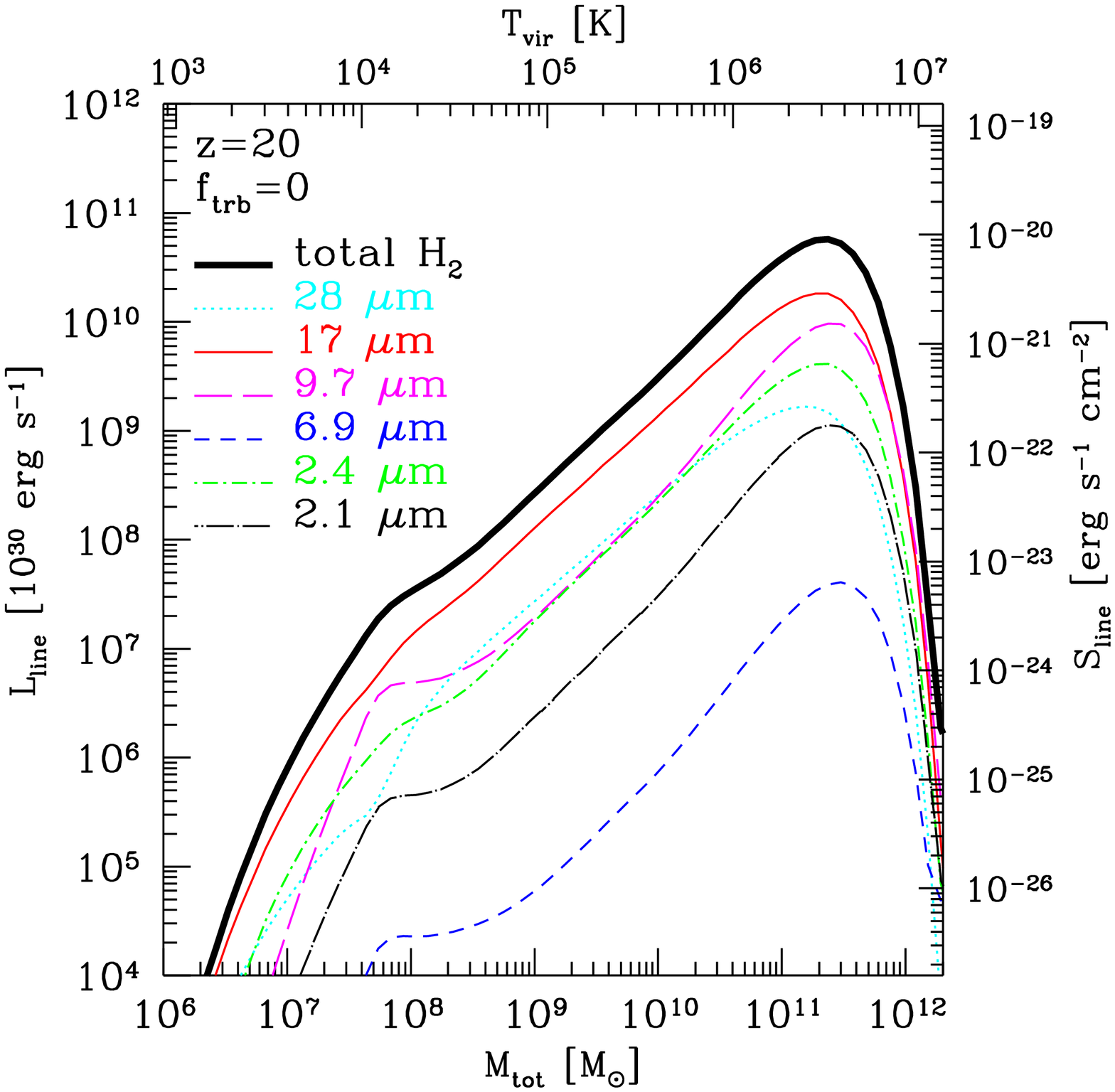}
\plotone{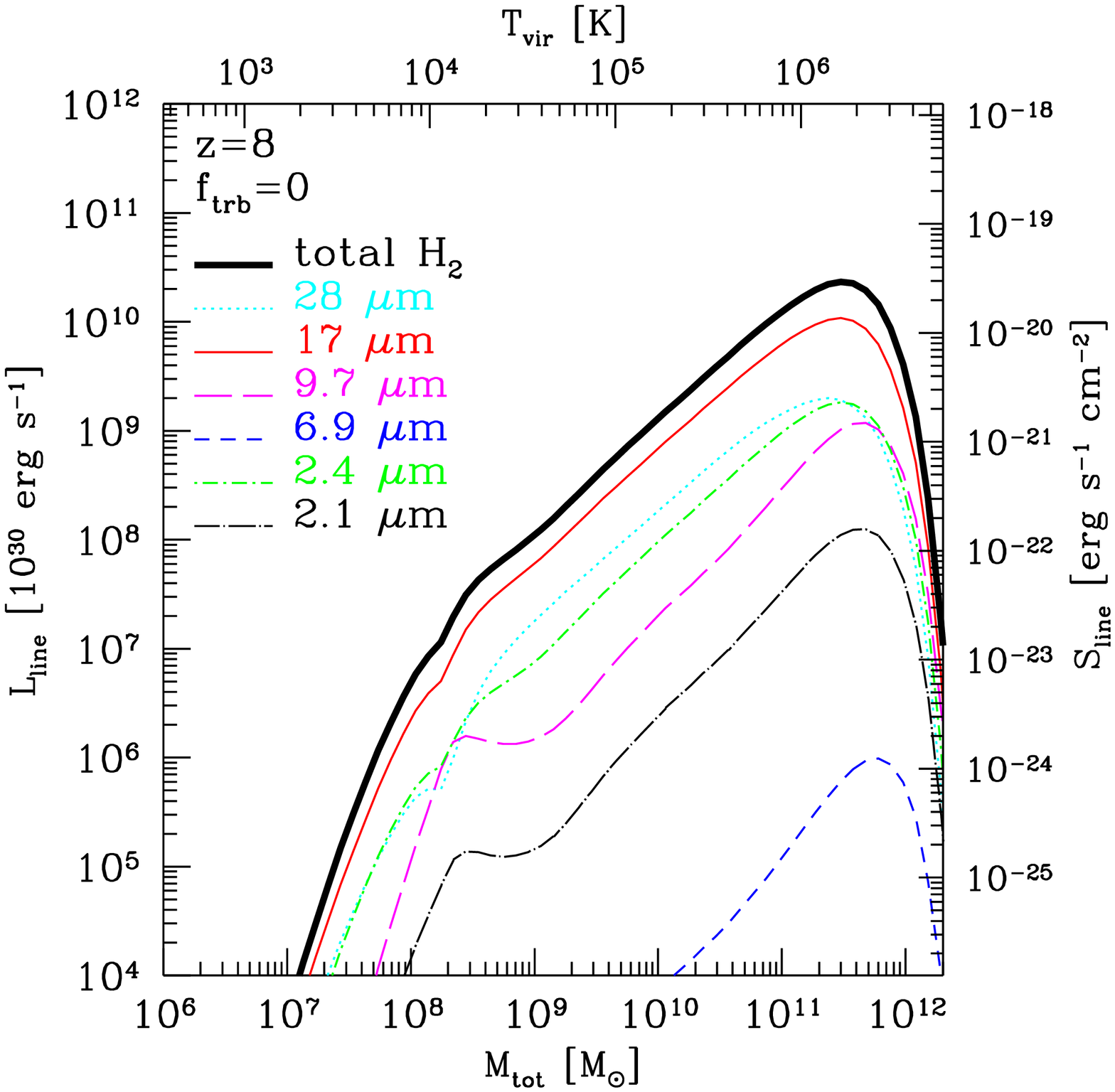}
\figcaption[f3a.eps;f3b.eps]
{The same as Figure 2 but for the dissipationless case
($f_{\rm trb}=0$).
\label{fig:f3}}

\plotone{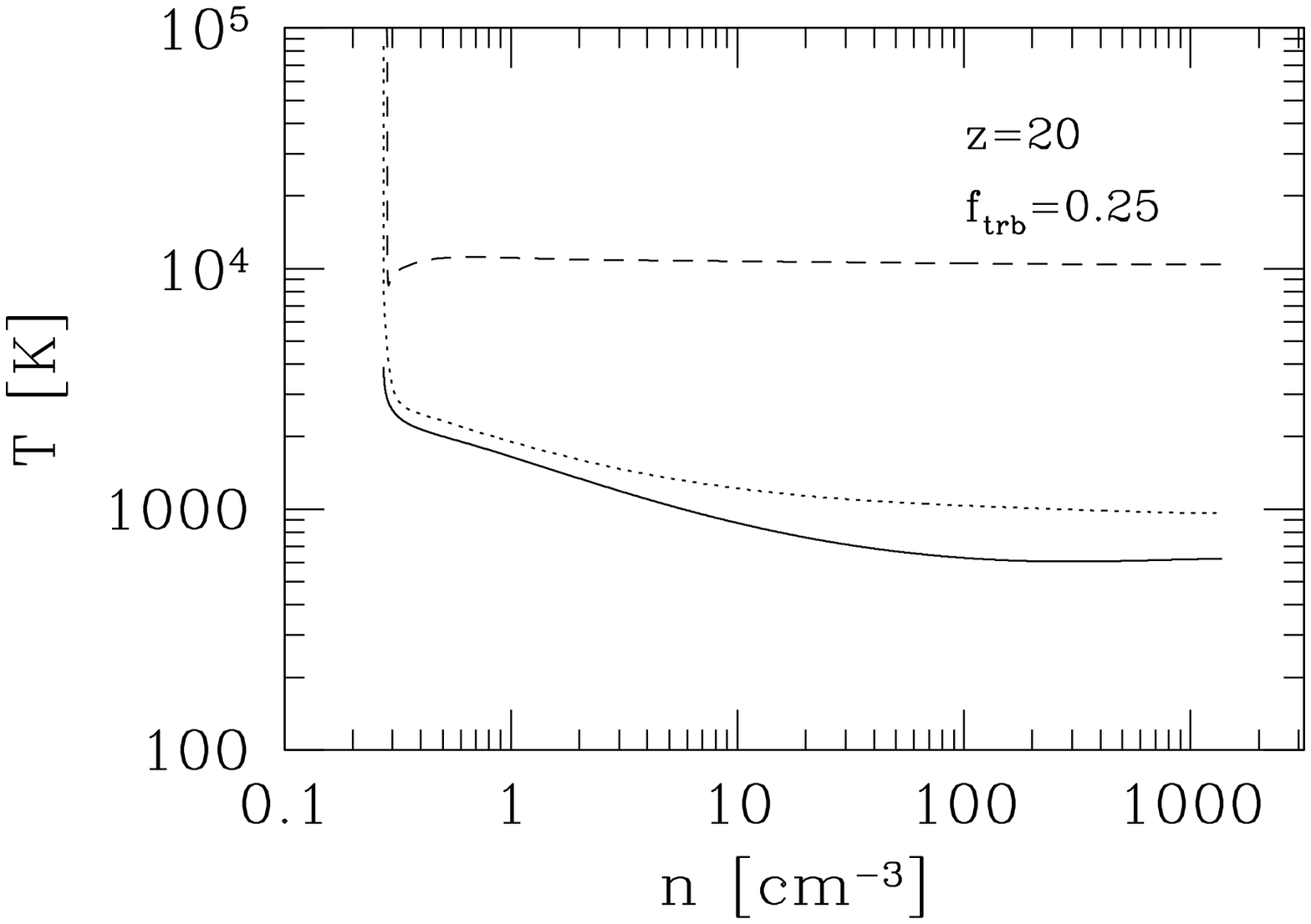}
\plotone{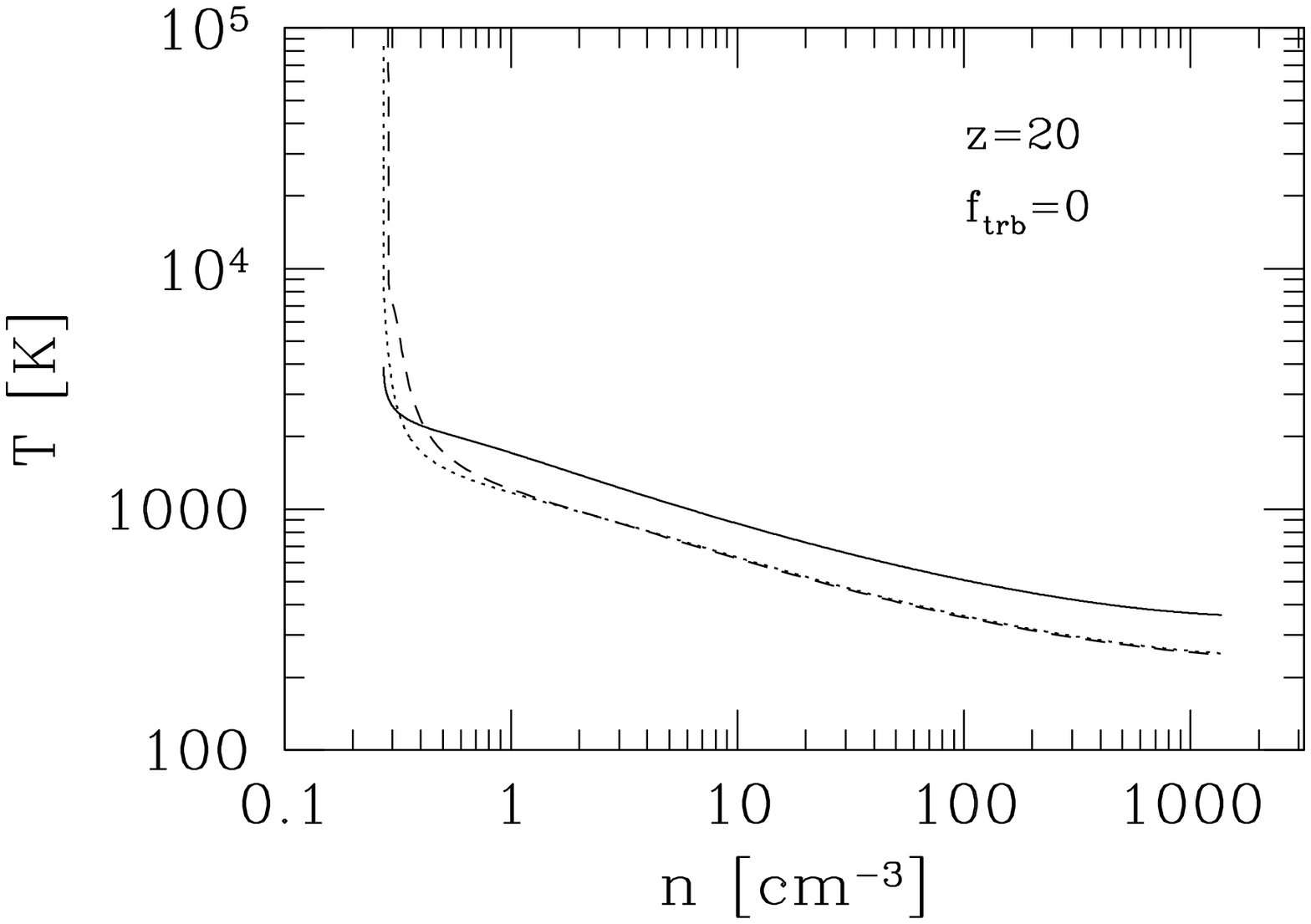}
\figcaption[f4a.eps;f4b.eps]
{The temperature evolution of protogalaxies virializing
at $z=20$ as a function of the H number density $n$ for
(a)the fiducial case ($f_{\rm trb}=0.25$), and (b)the dissipationless 
case ($f_{\rm trb}=0$).
Illustrated are the protogalaxies of $M_{\rm tot}=10^{7}$ (solid lines), 
$10^{9}$ (dotted lines),and $10^{11}M_{\rm sun}$ (dashed lines).
\label{fig:f4}}

\plotone{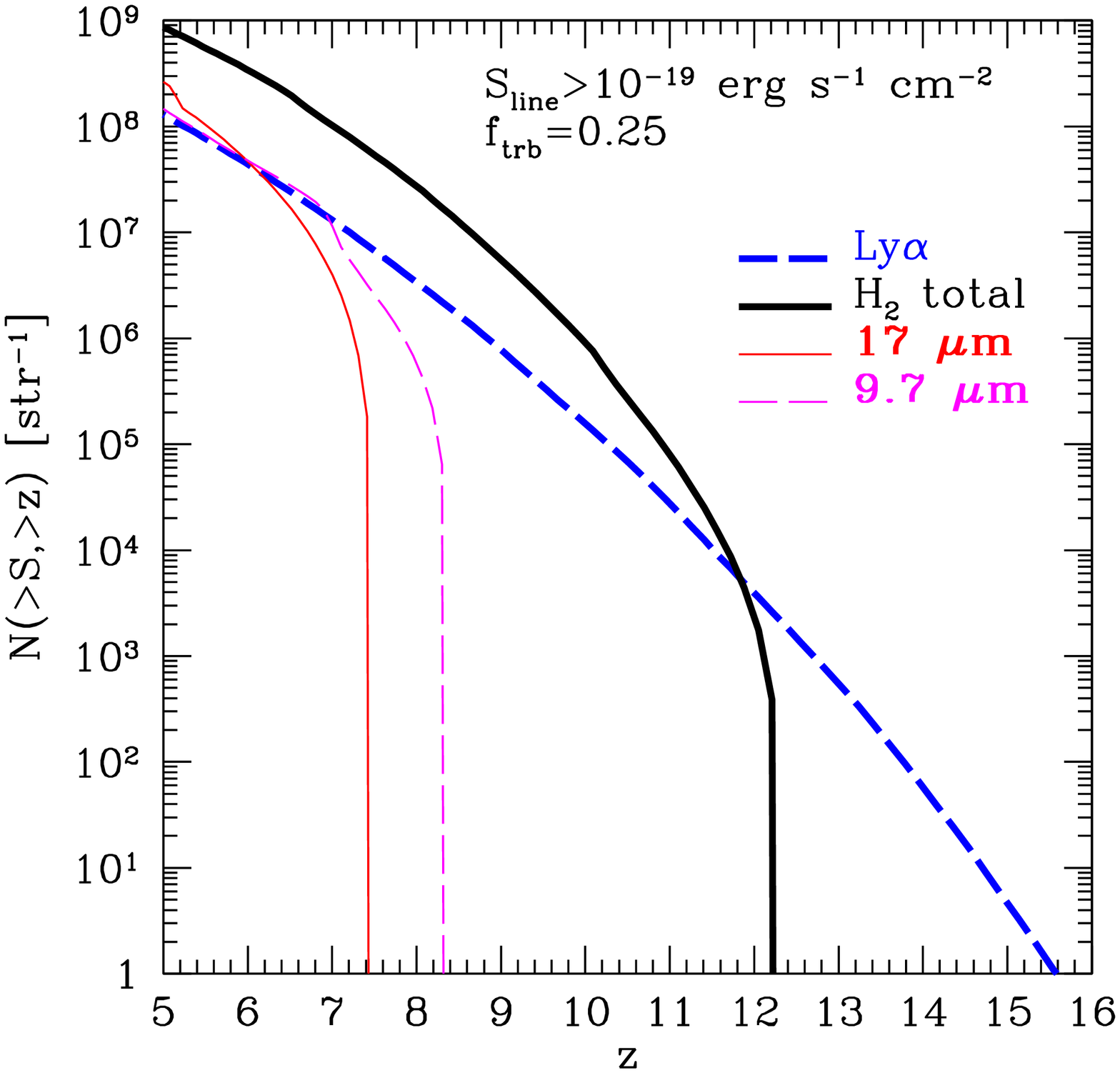}
\plotone{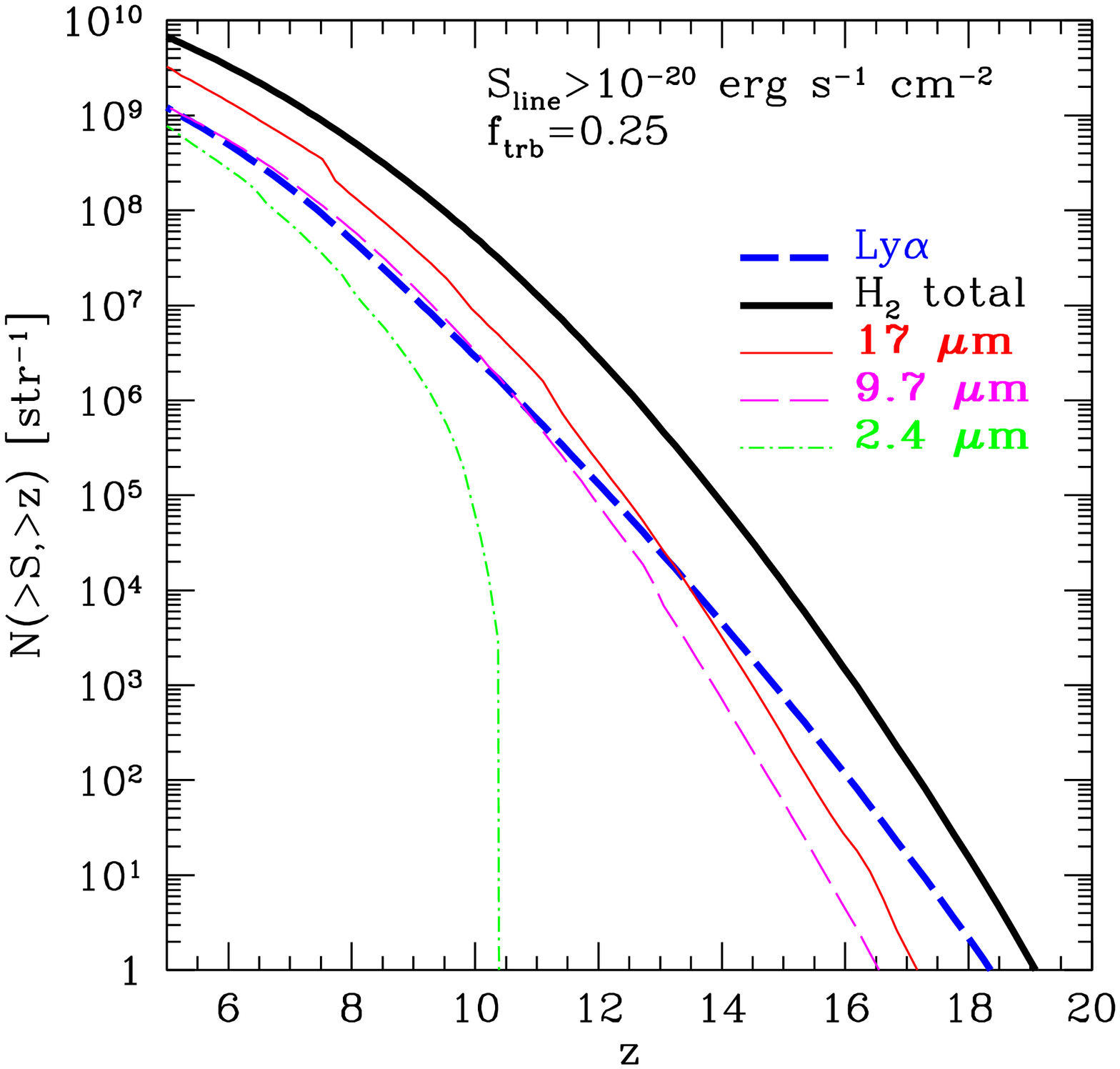}
\figcaption[f5.eps]{The number of sources below a redshift $z$ exceeding 
the limiting flux (a) $S_{\rm line}>10^{-19} ({\rm erg~s^{-1} cm^{-2}})$,
or (b) $S_{\rm line}>10^{-20} ({\rm erg~s^{-1} cm^{-2}})$.
The thick solid line shows the total flux of H$_2$ lines, and
the thick dashed line, the Ly$\alpha$ flux.
The flux of individual H$_2$ lines $17 \mu$m and $9.7 \mu$m are also shown 
in thin lines.
\label{fig:f5}}
 
\plotone{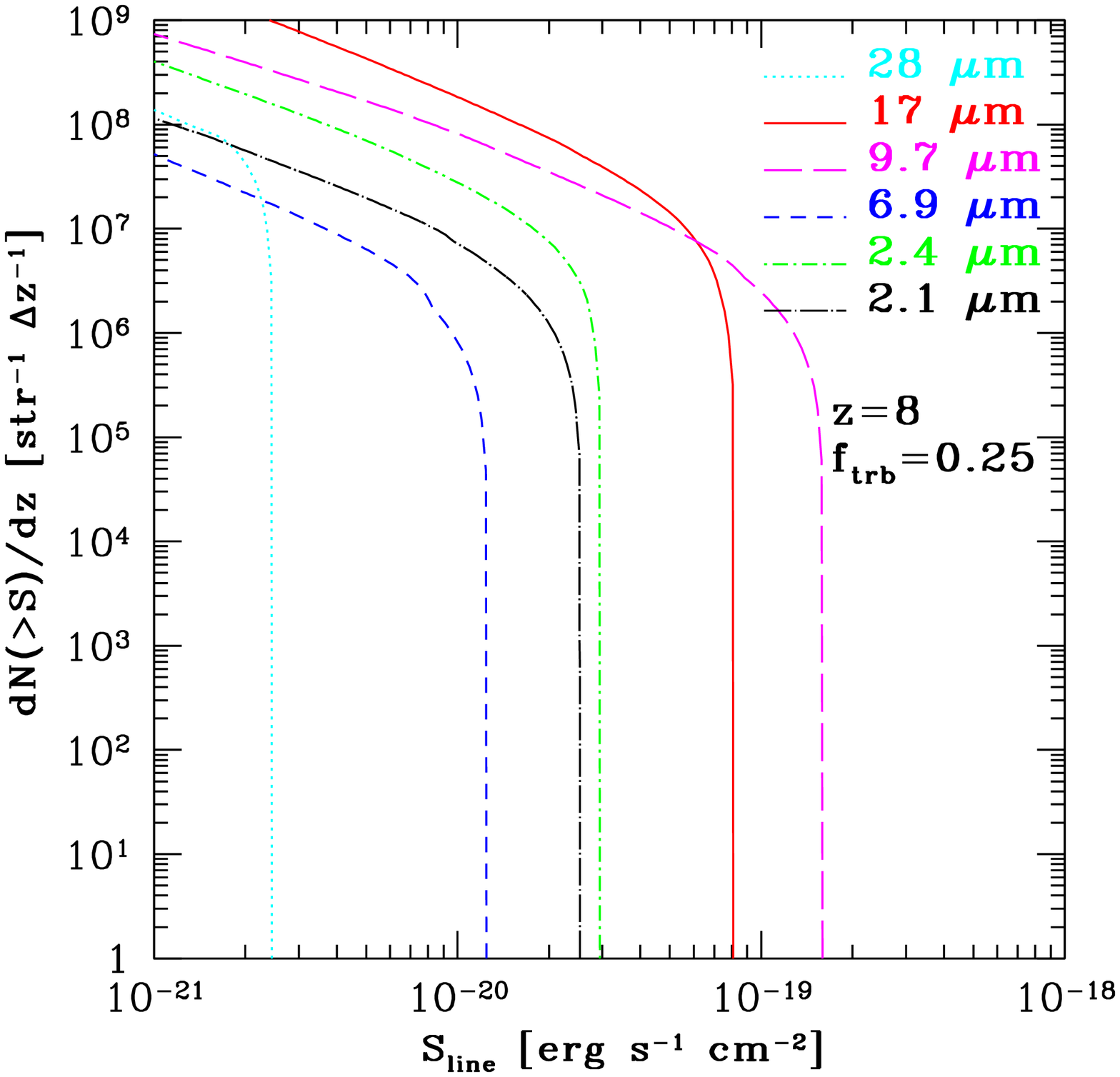}
\figcaption[f6.eps]{The number count of H$_2$ emitting sources 
as a function of the minimum observed flux $S_{\rm line}$ per unit redshift 
interval at $z=8$.
\label{fig:f6}}

\plotone{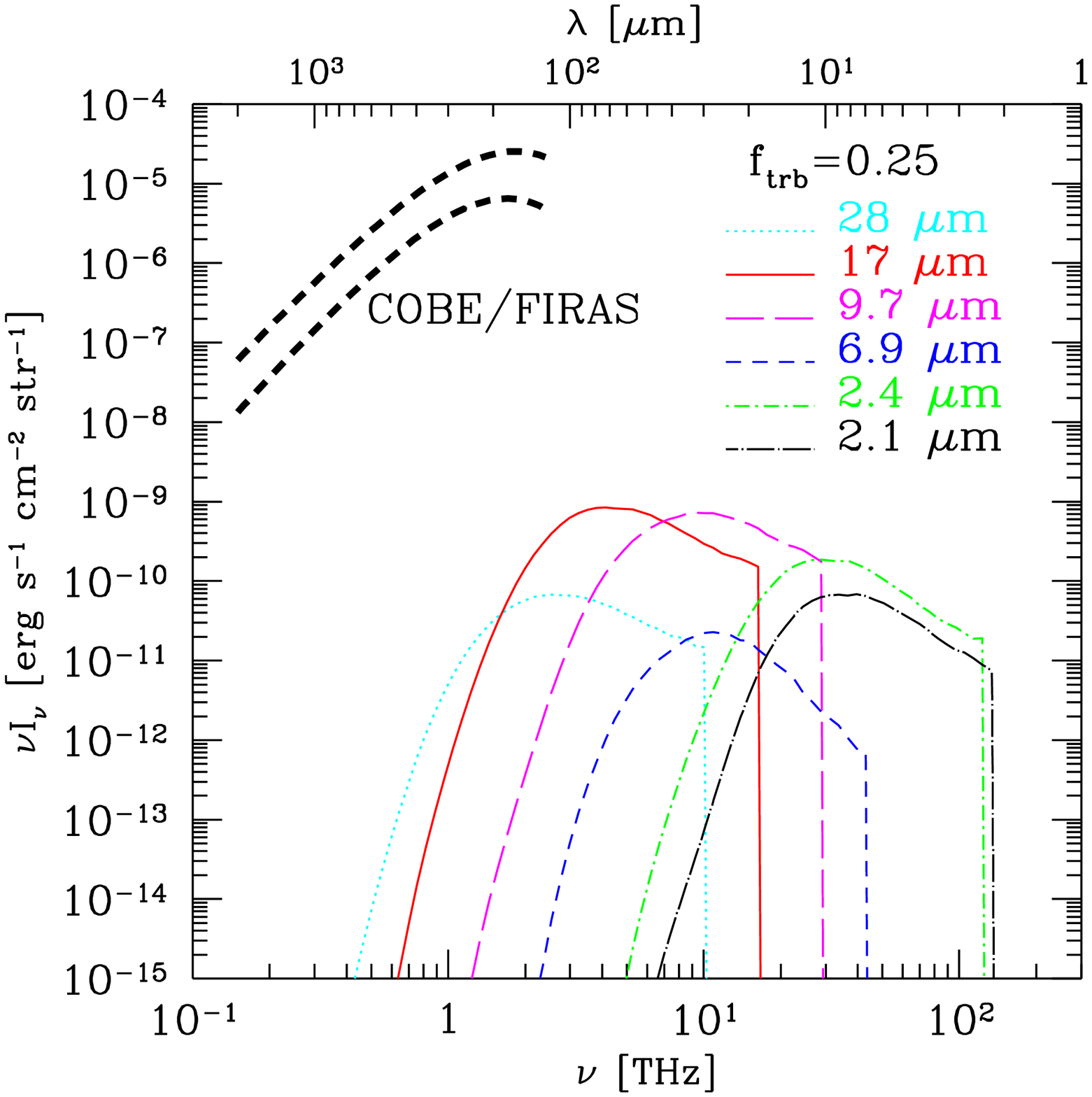}
\figcaption[f7.eps]{The contribution of H$_2$ cooling radiation to 
the cosmic background in the fiducial case ($f_{\rm trb}=0.25$). 
Also shown is the IR background detected by COBE/FIRAS. 
\label{fig:f7}}

\begin{table}
\begin{center}
\renewcommand{\arraystretch}{1.0}
\begin{tabular}{lcrllrlc} \hline
line & wavelength & luminosity & (erg/s) & & observed flux 
&(erg/cm$^{2}$/s) & \\
& ($\mu$m) & $10^{7}M_{\sun}$ & $10^{9}M_{\sun}$ & $10^{11}M_{\sun}$ &  
$10^{7}M_{\sun}$ & $10^{9}M_{\sun}$ & $10^{11}M_{\sun}$ \\
\hline \hline
0-0S(0) & 28.3 & 4.83E+34 & 5.49E+37 & 3.54E+38 & 7.94E-27 & 9.02E-24 & 
5.81E-23 \\
0-0S(1) & 17.1 & 3.42E+35 & 4.67E+38 & 2.50E+40 & 5.61E-26 & 7.66E-23 & 
4.11E-21 \\
0-0S(2) & 12.3 & 8.04E+34 & 1.27E+38 & 3.57E+40 & 1.32E-26 & 2.08E-23 & 
5.86E-21 \\
0-0S(3) & 9.69 & 2.48E+34 & 8.78E+37 & 1.34E+41 & 4.07E-27 & 1.44E-23 & 
2.20E-20 \\
0-0S(5) & 6.95 & 3.52E+31 & 1.11E+37 & 1.46E+39 & 5.78E-30 & 1.82E-24 & 
2.39E-22 \\
1-0O(3) & 2.81 & 6.85E+34 & 7.26E+37 & 4.98E+39 & 1.13E-26 & 1.19E-23 & 
8.18E-22 \\
2-1Q(1) & 2.55 & 1.33E+34 & 1.04E+37 & 2.48E+39 & 2.18E-27 & 1.72E-24 & 
4.07E-22 \\
1-0Q(1) & 2.41 & 7.99E+34 & 8.47E+37 & 5.81E+39 & 1.31E-26 & 1.39E-23 & 
9.54E-22 \\
1-0S(0) & 2.23 & 9.64E+33 & 1.12E+37 & 1.84E+39 & 1.58E-27 & 1.84E-24 & 
3.03E-22 \\
1-0S(1) & 2.12 & 4.49E+33 & 8.27E+36 & 7.37E+39 & 7.37E-28 & 1.36E-24 & 
1.21E-21 \\
1-0S(3) & 1.96 & 1.73E+31 & 2.83E+35 & 1.86E+39 & 2.84E-30 & 4.64E-26 & 
3.06E-22 \\
\hline 
total H$_2$&      & 7.63E+35 & 1.05E+39 & 2.66E+41 & 1.25E-25 & 1.72E-22 & 
4.37E-20 \\
\hline
H Ly$\alpha$ &   & 2.58E+26 & 6.95E+38 & 1.48E+42 & 4.23E-35 & 1.14E-22 & 
2.43E-19 \\
\hline
\end{tabular}
\end{center}
\caption{The H$_2$ luminosity and observed flux from forming galaxies of total 
mass $10^{7}M_{\sun}$, $10^{9}M_{\sun}$ and $10^{11}M_{\sun}$ at $z=20$.
The quantities of H Ly $\alpha$ are also presented.}
\end{table}
 
\begin{table}
\begin{center}
\renewcommand{\arraystretch}{1.0}
\begin{tabular}{lcrllrlc} \hline
line & wavelength & luminosity & (erg/s) & & observed flux 
&(erg/cm$^{2}$/s) & \\
& ($\mu$m) & $10^{7}M_{\sun}$ & $10^{9}M_{\sun}$ & $10^{11}M_{\sun}$ &  
$10^{7}M_{\sun}$ & $10^{9}M_{\sun}$ & $10^{11}M_{\sun}$ \\
\hline \hline
0-0S(0) & 28.3 & 9.34E+32 & 3.34E+37 & 1.27E+39 & 1.21E-27 & 4.31E-23 &
 1.64E-21 \\
0-0S(1) & 17.1 & 2.36E+33 & 2.57E+38 & 5.76E+40 & 3.04E-27 & 3.31E-22 &
 7.44E-20 \\
0-0S(2) & 12.3 & 2.33E+31 & 6.02E+37 & 5.05E+40 & 3.01E-29 & 7.77E-23 &
 6.51E-20 \\
0-0S(3) & 9.69 & 2.02E+30 & 3.99E+37 & 1.25E+41 & 2.60E-30 & 5.15E-23 &
 1.61E-19 \\
0-0S(5) & 6.95 & 8.02E+25 & 1.45E+36 & 9.87E+39 & 1.03E-34 & 1.87E-24 &
 1.27E-20 \\
1-0O(3) & 2.81 & 2.85E+32 & 2.45E+37 & 1.96E+40 & 3.68E-28 & 3.16E-23 &
 2.53E-20 \\
2-1Q(1) & 2.55 & 2.00E+31 & 3.05E+36 & 4.29E+39 & 2.58E-29 & 3.93E-24 &
 5.53E-21 \\
1-0Q(1) & 2.41 & 3.32E+32 & 2.85E+37 & 2.29E+40 & 4.29E-28 & 3.68E-23 &
 2.95E-20 \\
1-0S(0) & 2.23 & 2.08E+30 & 3.04E+36 & 6.20E+39 & 2.68E-30 & 3.92E-24 &
 8.00E-21 \\
1-0S(1) & 2.12 & 2.82E+29 & 2.51E+36 & 1.94E+40 & 3.63E-31 & 3.23E-24 &
 2.50E-20 \\
1-0S(3) & 1.96 & 9.26E+24 & 3.17E+34 & 2.74E+39 & 1.19E-35 & 4.09E-26 &
 3.53E-21 \\
\hline 
total H$_2$&   & 4.22E+33 & 4.89E+38 & 4.13E+41 & 5.44E-27 & 6.31E-22 &
 5.33E-19 \\
\hline
H Ly$\alpha$ & & 4.74E+08 & 3.84E+37 & 1.87E+41 & 6.11E-52 & 4.95E-23 &
 2.42E-19 \\
\hline
\end{tabular}
\end{center}
\caption{The same as Table 1 but for $z=8$}
\end{table}

\begin{thebibliography}{}
\bibitem[]{} Abel, T., Bryan, G. L., \& Norman, M. L. 2002, Science, 295, 93
\bibitem[]{} Anninos, P., Zhang, Y., Abel, T., Norman, M. L. 1997, 
NewA, 2, 209 
\bibitem[]{} Becker, R. H. et al. 2001, \aj, 122, 2850
\bibitem[]{} Bromm, V., Coppi, P. S., \& Larson, R. B. 2002, \apj, 564, 23
\bibitem[]{} Bryan, G. L., \& Norman, M. L. 1998, \apj, 495, 80
\bibitem[]{} Ciardi, B., \& Ferrara, A. 2001, \mnras, 324, 648
\bibitem[]{} Ciardi, B., \& Ferrara, A., Governato, F., \& Jenkins, A. 2000,
\mnras, 318, 1068
\bibitem[]{} Combes, F. 1999, in Proceedings of the ESO Workshop, 
Chemical Evolution from Zero to High Redshift, 
ed. J. R. Walsh, M. R. Rosa (Berlin: Splinger), 213
\bibitem[]{} Dyson, J. E., \& Williams, D. A. 1997, 
The Physics of the Interstellar Medium, 2nd edition (Bristol:IOP Publishing)
\bibitem[]{} Fall, S. M., \& Rees, M. J. 1985, \apj, 298, 18
\bibitem[]{} Fardal, M. A. et al. 2001, \apj, 562, 605
\bibitem[]{} Flower, D. R., \& Pineau des For\^{e}ts, G. 2001, \mnras,
	  323, 672
\bibitem[]{} Fixsen, D. J., Dwek, E., Mather, J. C., Bennett, C. L., 
Shafer, R. A.,  \apj, 508, 123
\bibitem[]{} Galli, D., \& Palla, F. 1998, \aap, 335, 403
\bibitem[]{} Haiman, Z., Rees, M. J., Loeb, A. 1997, \apj, 476, 458
\bibitem[]{} Haiman, Z., Spaans, M., \& Quataert, E. 2000, \apj, 537, L5
\bibitem[]{} Hu, E. M., Cowie, L. L., \& McMahon, R. G. 1998, \apj, 502, L99
\bibitem[]{} Izotov, Yu. I., \& Kolesnik, I. G. 1984, Soviet Astron., 28, 15
\bibitem[]{} Kamaya, H., \& Silk, J. 2002, \mnras, 332, 251
\bibitem[]{} Kitayama, T., Susa, H. Umemura, M., \& Ikeuchi, 2001, \mnras,
	  326, 1353 
\bibitem[]{} Kitayama, T., \& Suto, Y., 1996, \apj, 469, 480
\bibitem[]{} Kodaira, K., et al., 2003, \pasj, 55, L17
\bibitem[]{} Lacey, C. G., \& Cole, S., 1993, \mnras, 262, 627
\bibitem[]{} Maihara, T., et al. 2001, \pasj, 53, 25  
\bibitem[]{} Oh, S. P. \& Haiman, Z. 2002, \apj, 569, 558
\bibitem[]{} Ohta, K., et al. 1996, \nat, 382, 426
\bibitem[]{} Omont, A., et al. 1996, \nat, 382, 428
\bibitem[]{} Omukai, K. 2000, \apj, 534, 809
\bibitem[]{} Omukai, K. 2001, \apj, 546, 635
\bibitem[]{} Omukai, K. \& Nishi, R. 1998, \apj, 508, 141
\bibitem[]{} Omukai, K. \& Nishi, R. 1999, \apj, 518, 64
\bibitem[]{} Omukai, K. \& Palla, F. 2001 \apj, 561, L55
\bibitem[]{} Omukai, K. \& Palla, F. 2003 \apj, in press
\bibitem[]{} Partridge, R. B., \& Peebles, P. J. E. 1967, \apj, 147, 868
\bibitem[]{} Press, W. H., \& Schechter, P., 1974, \apj, 187, 425
\bibitem[]{} Rees, M. J., \& Ostriker, J. 1977, \mnras, 179, 541
\bibitem[]{} Ricotti, M., Gnedin, N. Y., \& Shull, J. M. 2002, \apj, 575, 49  
\bibitem[]{} Ripamonti, E., Haardt, F., Ferrara, A., \& Colpi, M. 2002,
\mnras, 334, 401
\bibitem[]{} Saslaw, W. C., \& Zipoy, D. 1967, \nat, 216, 976
\bibitem[]{} Shchekinov, Yu. 1991, \apss, 175, 57
\bibitem[]{} Shchekinov, Yu. A., \& \'{E}nt\'{e}l', M. B. 1985, 
Soviet Astron., 29, 491 
\bibitem[]{} Shibai, H., Takeuchi, T. T., Rengarajan, T. N., \& Hirashita, H.
2001, \pasj, 53, 589
\bibitem[]{} Silk, J. 1977, \apj, 211, 638
\bibitem[]{} Spaans, M., \& Silk, J. 1997, \apj, 488, L79
\bibitem[]{} Spergel, D., et al. 2003, \apj, submitted (astro-ph/0302209)
\bibitem[]{} Steidel, C. C., et al. 2000, \apj, 532, 170
\bibitem[]{} Susa, H., Uehara, H., Nishi, R., \& Yamada, M. 1998, 
Prog. Theor. Phys., 100, 63
\bibitem[]{} Taniguchi, Y., et al. 1997, \aap, 328, L9
\bibitem[]{} Williams, R. E., et al. 1996, \aj, 112, 1335
\end{thebibliography}
\end{document}